\documentclass[lettersize,journal]{IEEEtran}
\usepackage{amsmath,amsfonts}
\usepackage{algorithmic}
\usepackage{algorithm}
\usepackage{array}
\usepackage[caption=false,font=normalsize,labelfont=sf,textfont=sf]{subfig}
\usepackage{textcomp}
\usepackage{stfloats}
\usepackage{url}
\usepackage{verbatim}
\usepackage{graphicx}
\usepackage{cite}
\usepackage{epsfig}
\usepackage{fancyhdr}

\begin{document}

\title{Off-Network Communications For Future Railway Mobile Communication Systems: Challenges and Opportunities}

\author{Jiewen Hu, Gang Liu, \emph{Member, IEEE}, Yongbo Li, Zheng Ma, \emph{Member, IEEE}, Wei Wang, Chengchao Liang, \emph{Member, IEEE}, F. Richard Yu, \emph{Fellow, IEEE} and Pingzhi Fan, \emph{Fellow, IEEE}}



\maketitle

\thispagestyle{fancy}
\fancyhead{}
\lhead{\footnotesize \centering \copyright~2022 IEEE. Personal use of this material is permitted. Permission from IEEE must be obtained for all other uses, in any current or future media, including reprinting/republishing this material for advertising or promotional purposes, creating new collective works, for resale or redistribution to servers or lists, or reuse of any copyrighted component of this work in other works.}
\lfoot{}
\cfoot{}
\rfoot{}

\begin{abstract}

GSM-R is predicted to be obsoleted by 2030, and a suitable successor is needed. Defined by the International Union of Railways (UIC), the Future Railway Mobile Communication System (FRMCS) contains many future use cases with strict requirements. These use cases should ensure regular communication not only in network coverage but also uncovered scenarios. There is still a lack of standards on off-network communication in FRMCS, so this article focuses on off-network communication and intends to provide reference and direction for standardization. We first provide a comprehensive summary and analysis of off-network use cases in FRMCS. Then we give an overview of existing technologies (GSM-R, TETRA, DMR, LTE-V2X, and NR-V2X) that may support off-network communication. In addition, we simulate and evaluate the performance of existing technologies. Simulation results show that it is possible to satisfy the off-network communication requirements in FRMCS with enhancements based on LTE-V2X or NR-V2X. Finally, we give some future research directions to provide insights for industry and academia.
\end{abstract}


\section{Introduction}
Over the past 40 years, the world has rapidly evolved from 1G to 5G. The emergence of new technologies means the obsolescence of old technologies, and the Global System for Mobile Communications-Railway (GSM-R), which is based on 2G, is no exception. GSM-R is expected to become obsolete by 2030, making maintenance increasingly costly and complex. The International Union of Railways (UIC) group started looking for a replacement to GSM-R in 2015. In 2018 they established a structured outline named Future Railway Mobile Communication System (FRMCS). FRMCS is intended to replace the soon to be obsolete GSM-R and is also oriented to future railway applications, such as automated train operation, remote control, and future fully self-driving trains\cite{ref1}. This will require a more reliable and higher transmission rate communication system.

The issue of dedicated railway frequencies for FRMCS was raised as early as 2017, which boosted the development of FRMCS. In 2020, the European Conference of Postal and Telecommunications Administrations - Electronic Communications Committee (CEPT-ECC) decided to use the paired frequency bands 874.4-880.0 MHz, 919.4-925.0 MHz, and the unpaired frequency band 1900-1910 MHz for Railway Mobile Radio. In 2021, China planned to use 2.1 GHz as the operating band for future 5G-railway communication.

A core objective of FRMCS at the beginning was to standardize through the Third Generation Partnership Project (3GPP). Based on the FRMCS User Requirements Specification\cite{ref1} and Use Cases\cite{ref2}, the 3GPP studies and summarizes all the cases in FRMCS\cite{ref3}. In addition, the 3GPP further elaborates the off-network use case scenarios and refines their performance requirements in\cite{ref4}. Off-network communication refers to a direct communication mode between transmitter and receiver without passing the network, e.g., direct train-to-train (T2T) communication and direct train-to-ground (T2G) communication.
\begin{figure*}[!t]
\centering
\fbox{\includegraphics[width=7in]{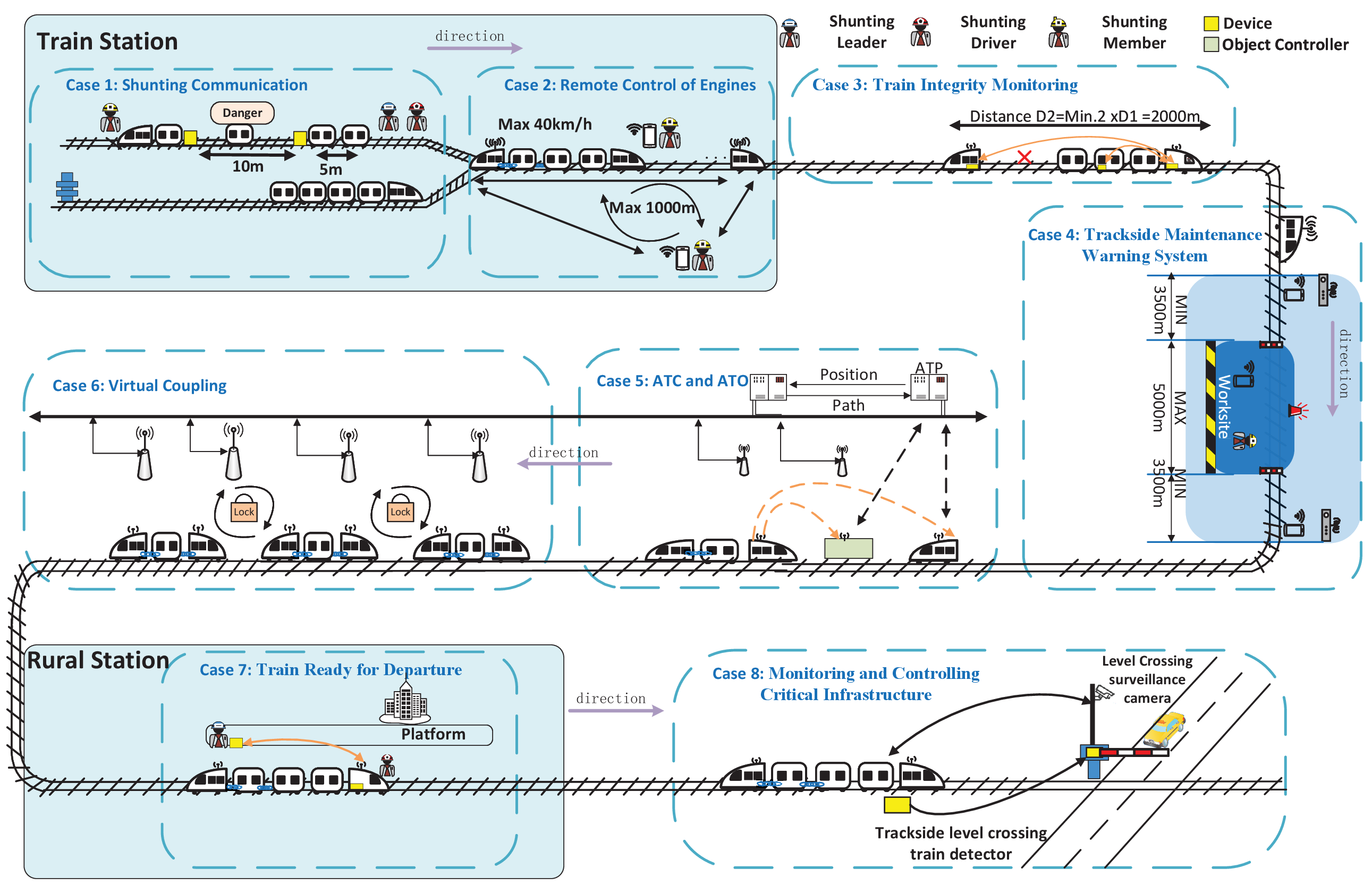}}
\caption{Use cases of off-network communication in FRMCS.}
\label{fig:1}
\end{figure*}

The research on railway off-network communication is in the very preliminary stage. The study in \cite{ref5} proposes a new train autonomous driving communication system based on 5G-T2T communication. The authors in \cite{ref6} modified Long Term Evolution Vehicle-to-Everything (LTE-V2X) for T2T communication and proposed a train-centric communication-based train control (CBTC) system. Most of these studies focus on automatic train control (ATC), but there are many scenarios of off-network communication in FRMCS, which have not been comprehensively studied. The article \cite{ref7} uses Terrestrial Trunked Radio (TETRA) technology for T2T communication and performs channel measurements at 450 MHz. The authors of \cite{ref8,ref9} take practical measurements of intelligent transport systems (ITS-G5) in T2T scenarios and propose a geometry-based stochastic channel model (GSCM) for T2T communication in open field environment. Although the existing works assume various existing technologies for railway off-network communication, there is still lack of a rigorous evaluation to judge whether they satisfy the requirements of off-network use cases in FRMCS.

This article first provides an overview of the off-network use cases in FRMCS, and introduces the existing technologies that may be used for future railway off-network communication, including GSM-R, TETRA, Digital Mobile Radio (DMR), LTE-V2X, and New Radio-V2X (NR-V2X). The performance of existing technologies is compared through simulation. Based on the simulation results, it is suggested that the enhancements on LTE-V2X or NR-V2X may satisfy the off-network communications requirements for FRMCS, which provides some potential directions and references for standardization. Finally, some possible research directions to improve transmission performance are given to provide insights for the future.

\section{use cases and requirements of off-network communication for railways}
This article mainly focuses on the off-network communication in FRMCS. Fig. \ref{fig:1} shows the use cases of off-network communication in FRMCS, and the performance requirements of off-network use cases are summarized in Table \ref{tab:1}\cite{ref4}.

\newcommand{\tabincell}[2]{\begin{tabular}{@{}#1@{}}#2\end{tabular}}
\begin{table}[!t]
\caption{Requirements of Off-network Communication Use Cases \label{tab:1}}
\centering
\scalebox{0.8}{
\begin{tabular}{|c|c|c|c|c|}
\hline
Scenario & \tabincell{c}{End-to-end\\latency} & Reliability & Data rate & \tabincell{c}{Communication\\range} \\
\hline
\tabincell{c}{ Shunting voice \\communication} & $\leq$ 100 ms & 99.9999 $\%$ & \tabincell{c}{100 kbps up\\to 300 kbps} & $\leq$ 1.5 km \\
\hline
\tabincell{c}{ Shunting data \\communication} & $\leq$ 500 ms & 99.9999 $\%$ & \tabincell{c}{10 kbps up\\to 500 kbps} & $\leq$ 1.5 km \\
\hline
\tabincell{c}{ Shunting video \\communication} & $\leq$ 100 ms & 99.9 $\%$ & 10 Mbps & $\leq$ 1.5 km \\
\hline
\tabincell{c}{ Remote control \\of engines data \\communication} & $\leq$ 10 ms & 99.9999 $\%$ & \tabincell{c}{100 kbps up\\to 1 Mbps} &  $\leq$ 1 km \\
\hline
\tabincell{c}{ Remote control \\of engines video \\communication} & $\leq$ 100 ms & 99.9 $\%$ & 10 Mbps & $\leq$ 1 km \\
\hline
\tabincell{c}{ Train integrity \\monitoring data \\communication} & $\leq$ 1 s & 99.9 $\%$ & \tabincell{c}{10 kbps up\\to 500 kbps} & $\leq$ 2 km \\
\hline
\tabincell{c}{ Trackside maintenance \\warning system \\communication} & $\leq$ 500 ms & 99.9999 $\%$ & \tabincell{c}{10 kbps up\\to 500 kbps} & $\geq$ 8.5 km \\
\hline
ATC and ATO & $\leq$ 100 ms & 99.99 $\%$ & $\leq$ 1 Mbps & $\leq$ 3 km \\
\hline
\tabincell{c}{ Virtual coupling \\  critical data\\communication} & $\leq$ 100 ms & 99.99 $\%$ & $\leq$ 1 Mbps & $\leq$ 3 km \\
\hline
\tabincell{c}{ Virtual coupling \\  very critical data\\communication} & $\leq$ 10 ms & 99.9999 $\%$ & $\leq$ 1 Mbps & $\leq$ 0.3 km \\
\hline
\tabincell{c}{ Train ready for \\ departure data\\communication} & $\leq$ 500 ms & 99.9 $\%$ & \tabincell{c}{10 kbps up\\to 500 kbps} & $\leq$ 1 km \\
\hline
\tabincell{c}{ Train ready for \\ departure video\\communication} & $\leq$ 100 ms & 99.9 $\%$ & 10 Mbps & $\leq$ 1 km \\
\hline
\tabincell{c}{ Monitoring and controlling \\  critical infrastructure\\video communication} & $\leq$ 100 ms & 99.9 $\%$ & $\leq$ 10 Mbps & $\leq$ 1 km \\
\hline
\end{tabular}
}
\end{table}

{\bf{Shunting Communication:}} Shunting movements include changing the locomotive of a train, coupling/uncoupling wagons, changing the order in which wagons are arranged in a train. All shunting movements are done within a station and require cooperation between shunting members via radio, which contains data, voice, and video communications over the network or off-network.

{\bf{Remote Control of Engines:}} The driver can remotely control the engine of the train via a ground-based system or an on-board system located at the other end of the engine. It is typically used for shunting operations in depots, shunting yards, or banking and should support off-network communication due to the uncertainty of the network.

{\bf{Train Integrity Monitoring:}} Accidental separation of train cabins is a hazardous event, and to avoid that situation, the driver can use this system to check the movement of the tail of the train while it is running. Even if the tail of the train is accidentally separated, the system still needs to maintain communication within a limited range after separation. Since the track is not always covered by the network when the train is running, the system should also be capable of off-network communication.

{\bf{Trackside Maintenance Warning System:}} Typically, the train is also running during trackside maintenance. To ensure the safety of the staff and the regular operation of the train, this system must promptly and accurately notify the staff of the upcoming train during the trackside maintenance period. Since the warning system is usually deployed temporarily and the network does not entirely cover the rail tracks, the system needs to have off-network communication capabilities.

{\bf{ATC and Autonomous Train Operation (ATO):}} This system is expected to be used primarily for urban rail transportation such as subways. Traditional train control system usually determines its movement by communicating with trackside devices, while this system allows trains to determine their movement autonomously based on direct communication between trains. For the safe operation of trains, even if the network is unavailable, the trains need to transmit speed, location, and other information in real-time, so this system needs to have the ability to communicate off the network.

{\bf{Virtual Coupling:}} This scenario is similar to vehicle platooning in V2X, where trains directly share control information (acceleration and braking, etc.) in real-time and combine it to control their trains. It can significantly reduce the distance between trains and increase the efficiency of railway transportation. Virtual coupling is that multiple trains in close distance move together as they are physically coupled. When the virtual coupling is formed, the distance between the trains is about 300 m (urban railway). Due to the high demand for latency, it is the only scenario that sets off-network communication as the default mode even when the network is available.

{\bf{Train Ready for Departure:}} This scenario is used to ensure that passengers can safely board the train and that the train can safely leave the platform. It consists primarily of driver-to-controller communication, conductor-to-driver communication, and platform camera video transmission. If the platform is in a remote area without a network, off-network communication is used.

{\bf{Monitoring and Controlling Critical Infrastructure:}} This system is designed mainly for video communication and is usually used for intersection monitoring, train detection, signals, indicators, etc. It should also be available in remote areas where there is no network coverage.

As shown in Table \ref{tab:1}, most of the off-network communication use cases have strict performance requirements, which require a minimum latency of 10 ms, a maximum data rate of 10 Mbps, and a maximum reliability requirement of 99.9999$\%$. In addition, the communication range is typically within 3 km, except for the trackside maintenance warning system, which needs to achieve reliability of 99.9999$\%$ over a communication range of more than 8.5 km and is extremely difficult to achieve for the existing technologies.

In general, the provisions of these off-network use cases are full of challenges. We need to answer the question that whether any existing technology can satisfy these stringent requirements. If no, we should find a successor or enhance on existing technologies.

\section{Overview of existing technologies}
This section provides a brief introduction to the existing technologies which may be used to support off-network communication. Among these technologies, GSM-R, TETRA, and DMR have been applied in some scenarios in the field of railway communication, while LTE-V2X and NR-V2X are technologies that can be used for off-network communication in the field of vehicular networking. Since there are many similarities between the two fields, it may be a good direction to consider the application of LTE-V2X and NR-V2X to railway off-network communication. The comparison of existing technologies is given in Table \ref{tab:2}.

\begin{table}[!t]
\caption{Comparison of existing technologies supporting off-network communication\label{tab:2}}
\centering
\scalebox{0.7}{
\begin{tabular}{|c|c|c|c|c|c|}
\hline
\tabincell{c}{Characteristics} & GSM-R \cite{ref10}& TETRA \cite{ref11}& DMR \cite{ref12}&  \tabincell{c}{LTE-V2X\\(sidelink)\\ \cite{ref13}} & \tabincell{c}{NR-V2X\\(sidelink)\\ \cite{ref14}}  \\
\hline
\tabincell{c}{Access\\scheme} & TDMA & TDMA & TDMA  & SC-FDMA & OFDMA \\
\hline
\tabincell{c}{Time slot\\duration}& 0.577 ms & 14.167 ms & 30 ms & 0.5 ms &  \tabincell{c}{1 / 0.5 /\\0.25 / 0.125 /\\ 0.0625 ms }\\
\hline
\tabincell{c}{Transmission\\power} & 30 dBm &  35 dBm & 30-46 dBm  & 23 dBm & 23 dBm \\
\hline
Frequency & \tabincell{c}{Uplink: \\876-915 MHz\\Downlink:\\ 921-960 MHz} & \tabincell{c}{380-400 MHz\\ 410-430 MHz\\ 450-470 MHz\\ 806-821 MHz\\851-866 MHz}& \tabincell{c}{30 MHz\\- 1 GHz} & \tabincell{c}{5855-5925 \\MHz (n47)} & \tabincell{c}{2570-2620 \\MHz (n38)\\5855-5925 \\MHz (n47)} \\
\hline
\tabincell{c}{Channel\\bandwidth} & 200 kHz & 25 kHz & 12.5 kHz & \tabincell{c}{1.4-20 \\ MHz} & \tabincell{c}{5-40 \\ MHz}\\
\hline
\tabincell{c}{Channel\\coding\\(data channel)}  & \tabincell{c}{Convolutional\\Code} & RCPC & \tabincell{c}{Trellis Code} & Turbo & LDPC \\
\hline
\tabincell{c}{Modulation\\scheme} & GMSK & $\pi/4$-DQPSK & 4FSK& \tabincell{c}{R14:\\QPSK\\16-QAM\\R15:\\64-QAM} & \tabincell{c}{QPSK\\16-QAM\\64-QAM\\256-QAM}  \\
\hline
\tabincell{c}{Peak \\transmission \\ rate\\(single channel )}  & \tabincell{c}{CSD: 9.6 kbps\\HSCSD: 14.4 kbps\\GPRS: 21.4 kbps} & 7.2 kbps & 4.8 kbps  & 30 Mbps &1 Gbps\\
\hline
\tabincell{c}{Operation \\Mode }  & \tabincell{c}{Infrastructure\\-based \\Mode only} & \tabincell{c}{Infrastructure\\-based \\/Direct Mode}  & \tabincell{c}{Infrastructure\\-based \\/Direct Mode}   & \tabincell{c}{Infrastructure\\-based \\/Direct Mode}  &\tabincell{c}{Infrastructure\\-based \\/Direct Mode}  \\
\hline
\tabincell{c}{Access \\protocol \\(Direct Mode) }  & -- & Slotted Aloha & Slotted Aloha  & \tabincell{c}{ Sensing-based\\SPS} &\tabincell{c}{ Sensing-based\\SPS}  \\
\hline
\tabincell{c}{Multi-antenna  \\supporting  }  &N& N & N  & Y &Y  \\
\hline
\end{tabular}
}
\end{table}
{\bf{GSM-R\cite{ref10}:}} GSM-R originated in Europe and is a radio standard dedicated to railway communications. It works near the 900 MHz frequency band, the access scheme is Time Division Multiple Access (TDMA), and the channel bandwidth is 200 kHz. It can achieve data rates of 9.6 kbps and 14.4 kbps per channel using Circuit Switched Data (CSD) mode and High Speed Circuit Switched Data (HSCSD) mode, respectively. In addition, it can achieve data rates of 21.4 kbps per channel when the General Packet Radio Service (GPRS) with Packet Switched Data (PSD) is used.

GSM-R defines four power levels to allow for different mobile devices, with a maximum of 8 W peak output power (1 W mean output power) and a minimum of 0.8 W output power. The data TCHs use convolutional code as the channel coding scheme, and the modulation scheme of GSM-R is Gaussian Minimum Shift Keying (GMSK) with parameter BT = 0.3, modulation rate of 270.83 kbit/s, and a Viterbi algorithm is used for demodulation.

{\bf{TETRA\cite{ref11}:}} TETRA is an open digital trunking system standard, mainly applicable to scenarios with emergency communication, daily command, scheduling, and other related mobile communication needs, and has vast application potential.

TETRA operates at 300 MHz to 1 GHz, the access method is TDMA, and the channel bandwidth shall be 25 kHz. The peak rate of the data traffic channel is 7.2 kbps. The average power of 0.56 W to 10 W is the common power range used by TETRA. The channel coding scheme for data traffic channels is Rate-Compatible Punctured Convolutional (RCPC) code. The modulation scheme is $\pi/4$-shifted Differential Quaternary Phase Shift Keying ($\pi/4$-DQPSK), and the random access protocol used by TETRA is based on slotted Aloha.

{\bf{DMR\cite{ref12}:}} DMR is a digital radio standard developed by the European Telecommunications Standards Institute (ETSI) specifically for professional mobile radio users. The DMR operates in part of the frequency range from 30 MHz to 1 GHz, with a TDMA access scheme and a channel bandwidth of 12.5 kHz.

The channel coding scheme of DMR is Forward Error Correction (FEC) codes, including Golay code, Hamming code, Block Product Turbo code (BPTC), and Trellis code. The Trellis code is used for data channels. The modulation scheme of DMR is 4-Frequency-shift keying (4FSK), and the average transmission power range is 1-40 W. The single-channel transmission rate can reach 4.8 kbps .

{\bf{LTE-V2X\cite{ref13}:}} LTE-V2X is a technology dedicated to vehicle communication, which can improve traffic efficiency and safety by exchanging information between vehicles and everything. LTE-V2X has two modes of transmission: Mode 3 and Mode 4. The Mode 4 is a direct mode, in which the vehicles reserve their resources autonomously through sidelink for communication without the support of network.

LTE-V2X operates in the 5.9 GHz (n47) band with a channel bandwidth of 1.4 MHz-20 MHz, and the access method is Single-Carrier Frequency Division Multiple Access (SC-FDMA). The channel coding scheme of LTE-V2X sidelink is turbo code. The modulation scheme is Quadrature Phase Shift Keying (QPSK) or 16-Quadrature Amplitude Modulation (16-QAM), and to support higher transmission rates, 3GPP added 64-QAM in release 15. The transmit power of LTE-V2X mobile devices are usually set to 23 dBm, supporting a transmission rate of 30 Mbps, and the transmission delay is generally in the range of 10-100 ms. LTE-V2X Mode 4 uses sensing-based semi-persistent scheduling (SPS) for resource reservation.

{\bf{NR-V2X\cite{ref14}:}}The 3GPP proposed NR-V2X in release 16 to supplement LTE-V2X in scenarios requiring high throughput, ultra-reliability, and low latency, and they will coexist and supplement each other in the future. Similar to LTE-V2X, NR-V2X also has two communication modes: Mode 1 and Mode 2. The Mode 2 is a direct communication mode with autonomous resource reservation through sidelink.

The frequency bands of NR include Frequency Range 1 (FR1) and Frequency Range 2 (FR2). The bands used by NR-V2X sidelink belong to FR1, which are 5.9 GHz (n47) and 2.5 GHz (n38), respectively. The channel coding scheme of NR-V2X sidelink is Low Density Parity Check (LDPC) code. Compared with LTE-V2X, the modulation scheme 256-QAM is added to support higher rate transmission. NR-V2X supports a channel bandwidth of 5-40 MHz, with a maximum transmission rate of 1 Gbps and an end-to-end latency as low as 3 ms\cite{ref15}. NR-V2X Mode 2 also uses the sensing-based SPS resource reservation scheme.

{\bf{Analysis of latency and transmission rate:}} GSM-R, TETRA, and DMR may have a latency of hundreds of milliseconds or even seconds, while LTE-V2X can achieve end-to-end latency of about 20 ms, and NR-V2X is even more advanced, reaching about 10 ms\cite{ref15}. As shown in Table \ref{tab:1}, most scenarios require end-to-end latency within 100 ms, and even some critical scenarios have a strict requirement of 10 ms. Therefore, the GSM-R, TETRA, and DMR may not satisfy the latency requirements of these use cases, while the LTE-V2X and NR-V2X may satisfy most of them.

In addition, GSM-R, TETRA, and DMR are narrowband technologies, and they have low data rates, which are insufficient for future off-network use cases of railways. On the contrary, the data rate performance of LTE-V2X and NR-V2X may meet the data rate requirements, which may reach 30 Mbps and 1 Gbps, respectively.

Besides end-to-end latency and data rate, the reliable communication distance of each existing technology is also a key performance indicator that still lacks a comprehensive evaluation. Therefore, the following section provides a detailed simulation analysis of the reliable communication distance.

\section{Reliable communication distance evaluation of existing technologies}
This section uses Matlab to evaluate the reliable communication distance of each existing technologies in an open field environment. The simulation sets up two trains, one in front of the other, with overhead line masts at regular intervals on the track, in addition to adjacent buildings, trees and walls. The channel model uses the GSCM consisting of line of sight (LOS) components and multipath components (MPCs). MPCs are composed of scattering components caused by overhead line mast, close buildings or trees, and reflection components caused by the ground or a wall. More detailed channel information and parameters can be found in \cite{ref9}.

The simulation evaluates the link-level performance of data channels for existing technologies considering the process of random data generation, channel coding, modulation, pass-through channel, demodulation, and decoding. Assuming that radio resources are sufficient and both trains can successfully reserve resources. The packet size is set to 256 octets \cite{ref4}, and each technology uses the modulation and channel coding schemes summarized in Table \ref{tab:2}, noting that LTE-V2X and NR-V2X use QPSK. In addition, the code rates of the channel coding schemes for all technologies are set to 1/3 for fairness.

\subsection{Reliable communication distance of different technologies with standard parameters}
\begin{figure}[!t]
\centering
\includegraphics[width=3.7in]{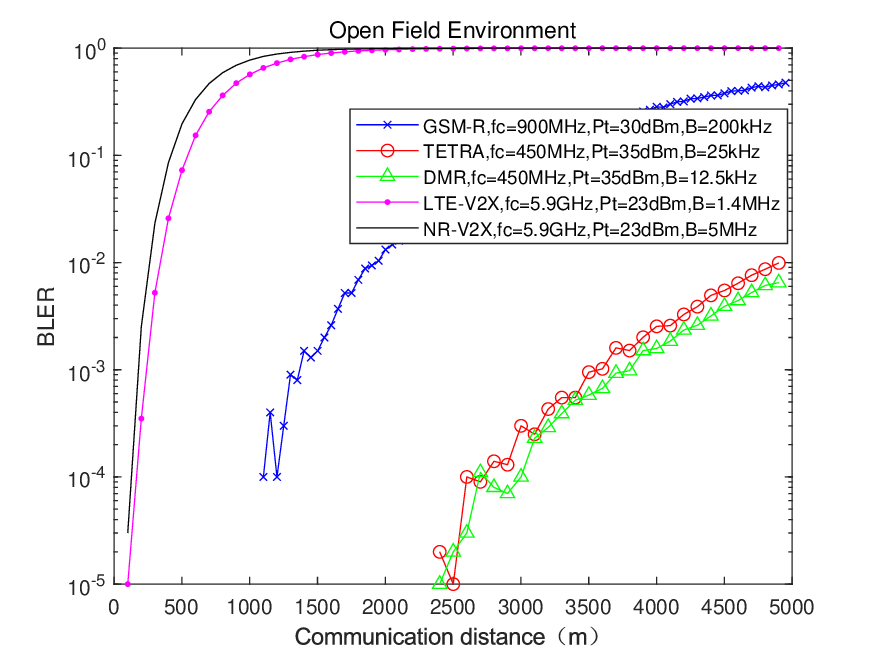}
\caption{BLER performance in open field environment with a single antenna and without retransmission when the physical layer parameters are specified in the standard.}
\label{fig:2}
\end{figure}

Fig. \ref{fig:2} shows the transport block error rate (BLER) performance of each technique at different communication distances. Each technique uses the frequency, transmission power and channel bandwidth given by the standard and shown in the legend. Combining the Table \ref{tab:1} and Fig. \ref{fig:2}, we can get the following conclusions:

Fig. \ref{fig:2} shows that LTE-V2X and NR-V2X have a reliable communication range of about 100 m. In contrast, GSM-R has a reliable transmission range of about 1 km, and TETRA and DMR have a reliable communication range of over 2.5 km. The reason is that LTE-V2X and NR-V2X have a low standard power of only 23 dBm and operate at 5.9 GHz. Higher frequency leads to more attenuation and therefore cannot satisfy the reliability and communication distance requirements of the use cases in Table \ref{tab:1}. So suitable transmission power and frequency are necessary.

According to the performance analysis at the end of Section III and the results in Fig. \ref{fig:2}, the existing technologies with the parameters given by the standard cannot fully satisfy the requirements of the use cases in Table \ref{tab:1}. In conclusion, to satisfy the requirements of off-network use cases in FRMCS, new technology or evolution of the existing technology is needed.
\begin{figure}[!t]
\centering
\includegraphics[width=3.7in]{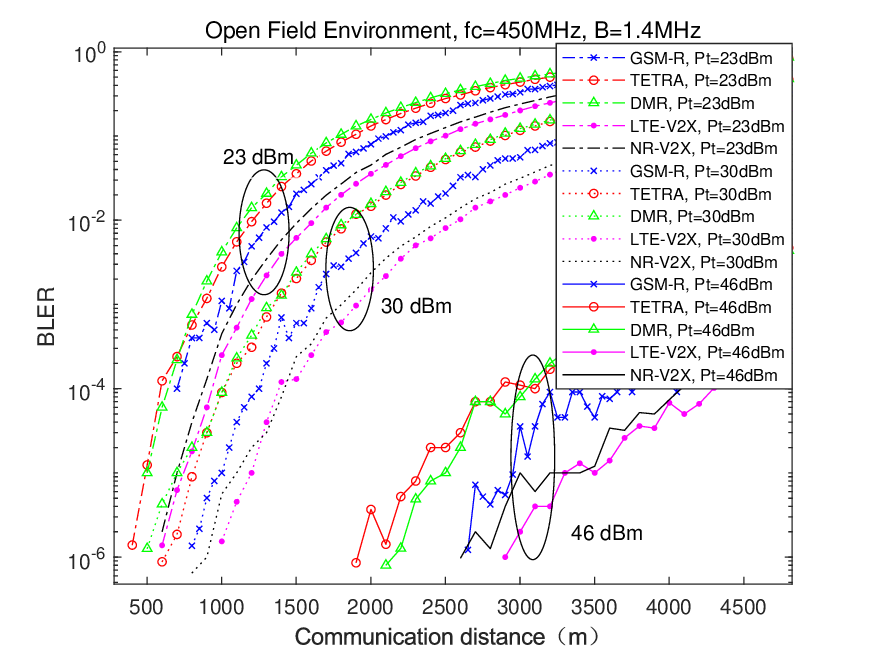}
\caption{BLER performance in open field environment with a single antenna and without retransmission when fc=450 MHz, B=1.4 MHz.}
\label{fig:3}
\end{figure}

\subsection{Impact of different parameters}
Based on the results of Fig. \ref{fig:2}, it is better to choose a lower frequency and an appropriate transmission power to achieve reliable transmission over long distances. Fig. \ref{fig:3} simulates the BLER performance of existing technologies at different communication distances by setting the frequency to 450 MHz, the bandwidth to 1.4 MHz, and the transmit power to 23 dBm, 30 dBm, and 46 dBm, respectively.

Fig. \ref{fig:2} shows the BLER performance of LTE-V2X and NR-V2X with 5.9 GHz is inferior to other low-frequency technologies. While Fig. \ref{fig:3} sets the frequency of LTE-V2X and NR-V2X to 450 MHz, their performance is significantly improved, even better than GSM-R, TETRA and DMR. The reason might be that the LDPC code used in NR-V2X and the turbo code used in LTE-V2X have more significant coding gain than the convolutional, RCPC and Trellis codes used in GSM-R, TETRA and DMR respectively, but also might introduce additional latency through coding delays, which will be further investigated in future work. Therefore, it is suggested to operate LTE-V2X or NR-V2X at a lower frequency to achieve long-distance transmission.

As shown in Fig. \ref{fig:3}, with the same settings of other parameters, the higher transmission power, the better performance is obtained. When the transmit power is increased to 46 dBm, which is usually used for base station, the existing technologies in open field environment can satisfy the reliability requirements of 99.99$\%$ for 3 km communication distance and 99.9999$\%$ for
1.5 km communication distance as mentioned in the Table \ref{tab:1}. But open field environment is only part of realistic scenarios, and there are also non-line-of-sight (NLOS) environments where greater fading is caused by obstruction. The performance of these techniques under NLOS environments would be worse than open field environment. Therefore even if the power is increased to 46 dBm, they may not satisfy the requirements of NLOS scenarios. So blindly increasing the transmission power is not feasible. Instead, if the existing LTE V2X or NR V2X is further enhanced, 30 dBm may be a more suitable power, which is now commonly used for trains communication.

According to the analysis in Fig. \ref{fig:3}, when using the 30 dBm power commonly used in trains, LTE-V2X and NR-V2X are insufficient to support the use cases in Table \ref{tab:1} in terms of reliable communication distance even with lower frequency. However, appropriate enhancement of LTE-V2X and NR-V2X may have the potential to fulfill these strict requirements. Therefore, besides adjusting the frequency and transmission power, we will further enhance them by adding Multiple Input Multiple Output (MIMO) and retransmission scheme in the next part and simulate them at the possible frequency bands of FRMCS to analyze their performance.


\begin{figure}[!t]
\centering
\includegraphics[width=3.7in]{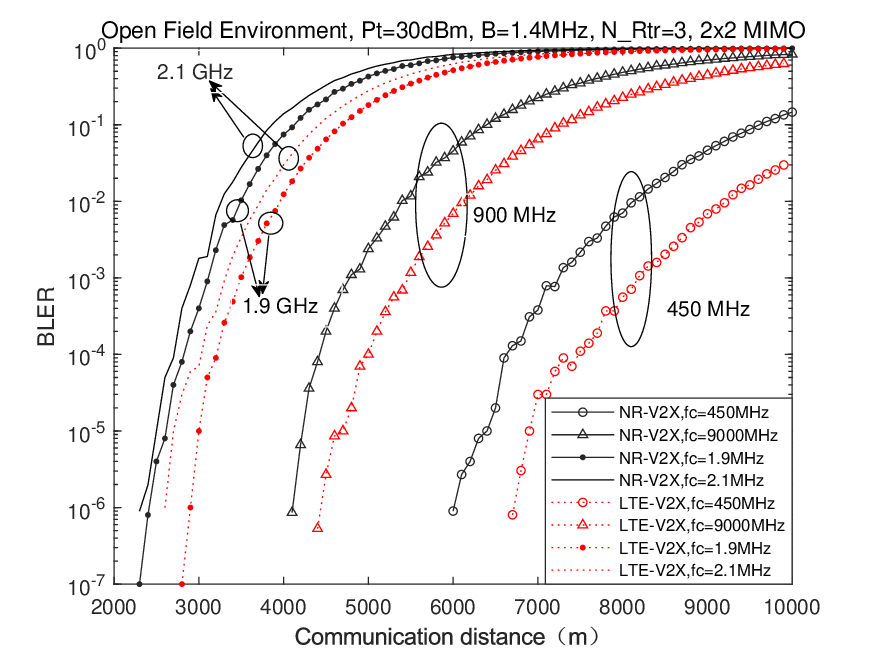}
\caption{BLER performance in open field environment with MIMO and retransmission when B=1.4 MHz, $P_t$=30.}
\label{fig:4}
\end{figure}
\subsection{Reliable communication distance of LTE-V2X and NR-V2X at FRMCS frequency band with MIMO and retransmission}
Fig. \ref{fig:4} simulates the LTE-V2X and NR-V2X technologies at the possible future FRMCS bands 900 MHz, 1.9 GHz and 2.1 GHz, and adds 450 MHz for comparison with Fig. \ref{fig:3}. The 2x2 MIMO is used to improve the performance further. Assuming that the antenna spacing is large enough so that antennas are not correlated with each other. The Minimum Mean Square Error (MMSE) based beamforming technique is used, and the 2x2 channel coefficients are generated by the extension of GSCM. In addition, since LTE-V2X only supports blind retransmission, Fig. \ref{fig:4} uses three blind retransmissions, and the transmission power is set to 30 dBm.

In Fig. \ref{fig:3}, when the transmission power is 30 dBm with the frequency of 450 MHz, LTE-V2X and NR-V2X can only achieve a reliability performance of 99.9999$\%$ less than 1 km communication distance. With the addition of MIMO and retransmission, Fig. \ref{fig:4} shows that they can achieve a reliability performance of 99.9999$\%$ over 6 km communication distance at 450 MHz in the open field environment, which is a significant improvement and is sufficient for most of the off-network communication use cases in Table \ref{tab:1}.

In addition, by adding MIMO and retransmission mechanism, the performance of LTE-V2X at 450 MHz, 900 MHz, 1.9 GHz and 2.1 GHz can satisfy the reliability requirements of 99.99$\%$ for 3 km communication distance and 99.9999$\%$ for 1.5 km communication distance as mentioned in the Table \ref{tab:1}. The NR-V2X also can satisfy these requirements at 450 MHz and 900 MHz. However, all performance evaluations are based on an open field environment. For NLOS scenarios, further evaluations will be done in future work.

In conclusion, enhancement of LTE-V2X and NR-V2X by adding MIMO and retransmission has the potential to satisfy all the off-network use cases requirements in Table \ref{tab:1} at 900 MHz and 450 MHz, except for the trackside maintenance warning system. For the consideration of policy permission, 900 MHz might be a more suitable frequency band for FRMCS off-network communication. The 900 MHz currently used by GSM-R may be refarmed for off-network communication in FRMCS.

\section{Opportunities and Challenges}
In this section, some potential research directions and challenges are presented to provide insight for future research.

{\bf{System Parameters Consideration:}} Frequency, transmission power, bandwidth, and modulation coding scheme significantly affect the performance of the system. A reasonable setting is required to meet the system requirements.

{\bf{Multi-air Interface and Multi-antenna:}} When the off-network communication mode is introduced, the placement and orientation of the antenna cannot be well optimized for train-to-ground communication, and at the same time for train-to-train communication. When performing mode switching, adjusting antenna will inevitably cause communication interruptions. Similar to the Uu interface (on-network mode) and PC5 (sidelink) interface (off-network mode) in LTE-V2X/NR-V2X, adding a new air interface for railway off-network mode is a good solution (the two air interfaces are independent of each other and can operate simultaneously). The existing Uu-PC5 seamless switching optimization scheme has been relatively mature, which has great reference value for future research on mode switching in railway.

Multiple antennas need a trade-off between diversity and multiplexing, and the high-speed movement of trains can make channel estimation inaccurate. Another key challenge is how to optimize and deploy the antennas.

{\bf{MAC Protocol Design:}} If V2X is used for FRMCS off-network communication, the higher transmit power of the PC5 interface will make coexistence with cellular link (Uu) in the same band challenging. In addition, since there is no centralized control center for off-network communication, suppressing interference from other users in the same frequency band and avoiding collisions when reserving the channel are also key challenges. The power control algorithm, channel division protocol, and random access protocol of the MAC layer can be used to avoid these problems. Therefore, the MAC protocols dedicated to the railway environment need to be designed for FRMCS.

{\bf{Channel Measurement:}} An accurate channel measurement can help better understand the signal propagation conditions, which can benefit system design and parameter setting. There have been already works on T2T channel measurement, such as \cite{ref7,ref8,ref9}. However, the future railway off-network communication scenarios are diverse, and there might be several different potential frequencies. Hence, more channel measurements can be done for different scenarios with different frequencies.

{\bf{Off-network Communication Assisted by Relay or Satellite Communications:}} Some extreme scenarios, such as the trackside maintenance warning system, which requires reliability of 99.9999$\%$ within a communication range of 8.5 km, need some other technologies to assist communication. Existing technologies cannot achieve the long-range reliable communication requirements of trackside maintenance warning system with only one-hop transmission. Relay technology for multi-hop communication can be adopted to improve communication distance and reliability. In addition, satellite communication technology can also be used to support this use case.

{\bf{Mission Critical Framework (MCX):}} The FRMCS system builds upon the 3GPP MCX framework, which complements the transport technology by functions for authentication, functional addressing, etc. A key challenge is that the MCX framework is typically centralized and has not been used for off-network communications as of today. 3GPP expects to use ProSe direct communication to provide MCX, but lacks a set of standards and requires a lot of future research.

{\bf{System-level Simulation:}} This article performs link-level simulations of the data channel for existing technologies. In the future, more comprehensive system-level simulations are needed to evaluate the overhead and latency, including evaluation of the MAC layer as well as control channel overhead, etc.

{\bf{Coexistence of Multiple Technologies:}} Most of the existing railway communications are based on GSM-R, and it will be a long process to complete the transition from GSM-R to its successor (e.g., LTE-R or 5G-R), so they will coexist for a long time in the future. The coexistence of multiple technologies is also a challenge in terms of interference and allocation of bandwidth resources.

\section{Conclusion}
This article briefly introduces off-network use cases in FRMCS and summarizes the performance requirements of these use cases. Then we provide an overview of existing technologies (GSM-R, TETRA, DMR, LTE-V2X and NR-V2X) that may support off-network communication and discuss their physical layer characteristics. In addition, we simulate and evaluate the data channel of existing technologies and compare their performance with the requirements of off-network use cases in FRMCS. According to the performance analysis of different technologies, it is challenging for the existing technologies to fully satisfy the requirements of these cases. However, with the help of MIMO and retransmission scheme, we point out that enhancement on LTE-V2X or NR-V2X operating at 900 MHz has the potential to satisfy the requirements of off-network use cases in FRMCS. Finally, some future research directions are suggested, including system parameters consideration, multi-air interface and multi-antenna, channel measurement, off-network communication assisted by relay or satellite communications, mission critical framework, MAC protocol design, system-level simulation and coexistence of multiple technologies.

\section*{Acknowledgments}
This work is jointly supported by NSFC project (grant No.61971359), Chongqing Municipal Key Laboratory of Institutions of Higher Education (grant No. cqupt-mct-202104), Fundamental Research Funds for the Central Universities, Sichuan Science and Technology Project (grant no. 2021YFQ0053) and State Key Laboratory of Rail Transit Engineering Informatization (FSDI).


\section*{Biographies}
Jiewen Hu is currently pursuing the Ph.D. degree at the Department of Information and Communication Engineering.

\ \notag\

Gang Liu [M'15] is currently an associate professor at School of Information Science and Technology, Southwest Jiaotong University, China.

\ \notag\

Yongbo Li is currently pursuing the Ph.D. degree at the Department of Information and Communication Engineering.

\ \notag\

Zheng Ma [M'07] is currently a professor at School of Information Science and Technology, Southwest Jiaotong University, China.

\ \notag\

Wei Wang is currently a senior engineer at State Key Laboratory of Rail Transit Engineering Informatization (FSDI), China.

\ \notag\

Chengchao Liang [S'15, M'17] He is currently a Professor with the School of Communication and Information Engineering, CQUPT.

\ \notag\

F. Richard Yu [S'00, M'04, SM'08, F'18] is currently a Professor at Carleton University, Ottawa, ON, Canada.

\ \notag\

Pingzhi Fan [M'93, SM'99, F'15] is currently a distinguished professor at Southwest Jiaotong University, China.
\newpage


\vfill

\end{document}